\documentclass[amsmath,floatfix,twocolumn,superscriptaddress,prx,longbibliography]{revtex4-1}
\usepackage{amssymb,amsmath}
\usepackage{graphicx,array}
\usepackage{dcolumn}
\usepackage{bm}
\usepackage{float}
\usepackage{caption}
\usepackage{setspace}
\usepackage{xkeyval}
\usepackage{rotating}

\begin{document}

\title{Simulation of the electrocaloric effect based on first-principles methods}

\author{Claudio Cazorla}
\email{c.cazorla@unsw.edu.au}
\affiliation{School of Materials Science and Engineering, UNSW Australia, Sydney NSW 2052, Australia}

\begin{abstract}
Due to critical environmental and technological issues, there is a pressing need to
switch from current refrigeration methods based on compression of gases to novel
solid-state cooling technologies. Solid-state cooling is based on the thermal response
of materials to external magnetic, electric, or mechanic fields, the so-called caloric
effect. The electrocaloric (EC) effect, which is caused by electric fields and
typically occurs in polar materials, is particularly promising from a technological
point of view owing to its good scalability and natural implementation in circuitry.
Simulation of EC effects represents an efficient and physically insightful strategy
for advancing the field of solid-state cooling by complementing, and in some cases
guiding, experiments. Theoretical estimation of EC effects can be achieved with
different approaches ranging from computationally inexpensive but physically
insightful phenomenological free-energy models to computationally very demanding
and quantitatively accurate first-principles methods. In this Chapter, we review 
EC simulation approaches that rely on first-principles methods. In this category, 
we include {\em ab initio} quasi-harmonic methods, bond-valence and classical interatomic 
potentials and effective Hamiltonians. In analogy to the experiments, these simulation 
approaches can be used to estimate EC effects either directly or indirectly and we 
review here well-established protocols that can be followed for each case. The Chapter 
finalises with a collection of representative examples in which first-principles based 
approaches have been used to predict and understand original EC effects.
\end{abstract}

\maketitle

\tableofcontents

\section{Introduction}
\label{sec:intro}
Polar materials exhibit a net electric dipole, also called polarization, which can be
modified by temperature or an electric field. From a crystallographic point of view, 
polar materials are characterized by non-centrosymmetric atomic structures that lack
inversion symmetry. Oxide perovskites like BaTiO$_{3}$ and PbTiO$_{3}$ are archetypal
polar materials in which their electric polarization is strongly coupled with their 
structural degrees of freedom [\onlinecite{zhong94}]. Magnetism may coexist with polar order 
in some of these compounds, the so-called multiferroics (e.g., BiFeO$_{3}$ and BiCoO$_{3}$
[\onlinecite{cazorla13,cazorla17,heo17}]), which may add further functionality to this class 
of materials [\onlinecite{spaldin19}]. Oxide perovskites can be synthesized in a wide 
variety of forms and sizes, like ceramics, thin films, nanocrystals and nanowires, by 
using well-established synthesis methods (e.g., solid-state reactions and chemical vapour and 
pulsed laser deposition techniques). Thus, owing to their unique functionality and morphological 
versatility, polar materials are found in a large number of technological applications 
related to the fields of information storage, electronic devices, and energy conversion 
[\onlinecite{rappe13,cazorla19a,catalan16}]. 

The isothermal entropy changes associated with fluctuations in the electric, magnetic and structural 
properties of polar materials can be large, $|\Delta S| \sim 100$~kJK$^{-1}$kg$^{-1}$, hence they may 
render sizable caloric effects (i.e., adiabatic temperature changes of $|\Delta T| \sim 10$~K). 
Solid-state cooling is an environmentally friendly, highly energy-efficient, and highly scalable 
technology that exploits caloric effects in materials for refrigeration purposes 
[\onlinecite{cazorla19b,moya14,cazorla17b,cazorla16,scott11,pecharsky97,bonnot08}]. By applying 
external fields on caloric materials it is possible to achieve reversible temperature shifts 
that can be integrated in cooling cycles and do not involve the use of greenhouse 
gases. Due to their natural implementation in circuitry, polar materials are specially well 
suited for solid-state cooling applications based on electrocaloric (EC) effects, which are 
driven by electric fields (see, for instance, works [\onlinecite{shirsath20,lisenkov09,wang20,
herchig15,cazorla15b}]).         
 
The magnitude of EC effects can be estimated accurately with theoretical and atomistic simulation 
methods. Computer simulations can be used to rationalize the origins of experimentally observed EC 
phenomena since the materials and phase transitions of interest can be accessed at the atomic scale  
under controlled conditions. Moreover, from a resources point of view computer simulations are 
inexpensive. For instance, by using open-source software and modest computer resources one already 
can simulate complex EC effects and assess the magnitude of the accompanying isothermal entropy and 
adiabatic temperature shifts. Thus, modeling of EC effects can be done systematically in order to 
complement and also guide the experiments.   

\begin{figure*}[t]
\centerline
        {\includegraphics[width=0.90\linewidth]{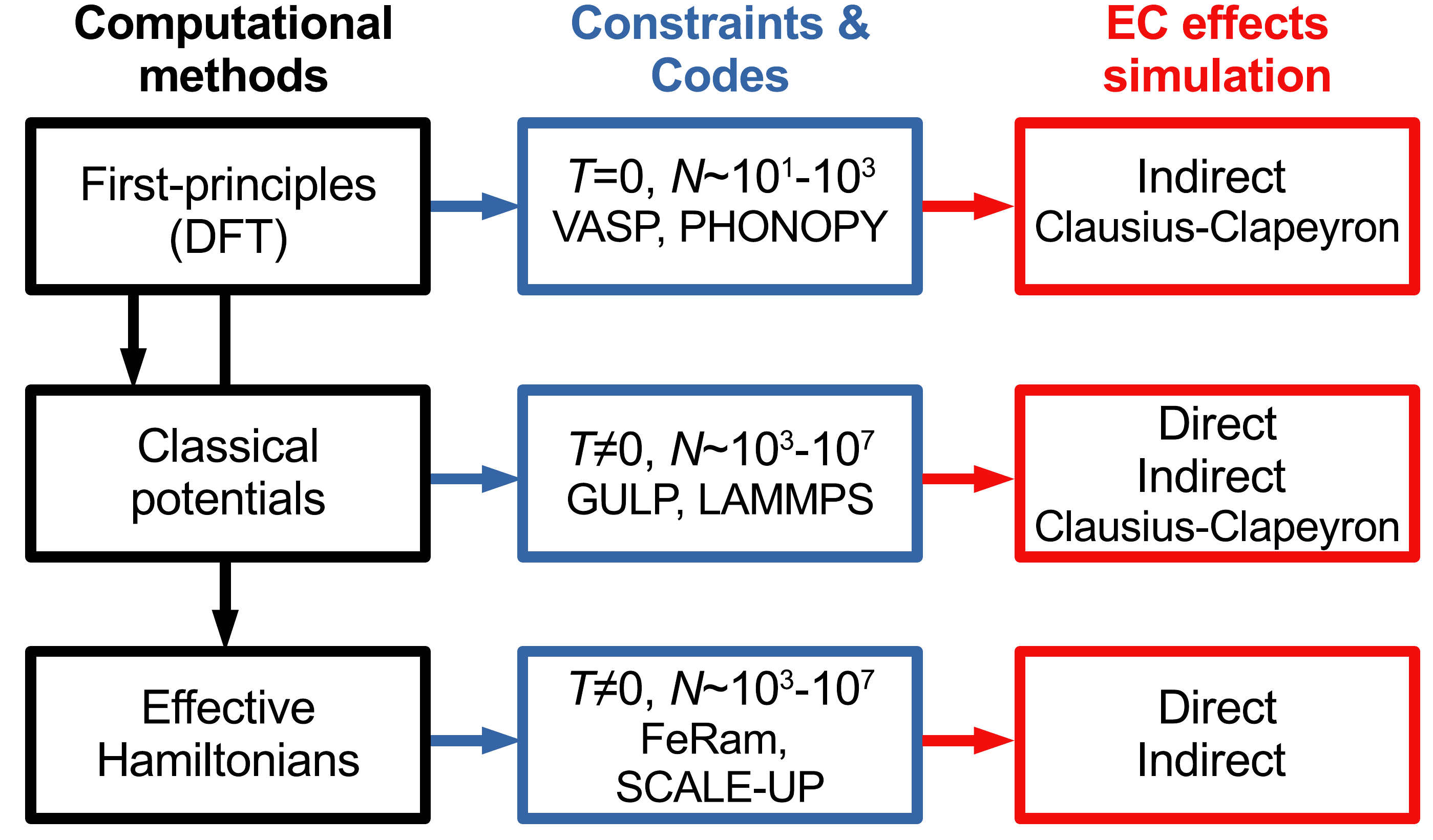}}
\caption{Overview of computational approaches based on first-principles methods that can be used
         to estimate EC effects. The typical system sizes that can be simulated with them, 
         $N$, along with some software codes in which they have been implemented, and the usual 
         modalities of EC simulation that can be done with them are indicated. VASP [\onlinecite{vasp}],
         PHONOPY [\onlinecite{phonopy}], GULP [\onlinecite{gulp}], LAMMPS [\onlinecite{lammps}], 
         FeRam [\onlinecite{feram}], and SCALE-UP [\onlinecite{scaleup}] stand for software 
         codes.}
\label{fig1}
\end{figure*}

The reliability of computer simulations, however, depends strongly on the simplifications made 
on the adopted structural and interaction models. Typically, increasing the reliability of the 
structurals model comes at the expense of reducing the accuracy in the description of the interatomic 
forces (due to practical limitations, see Fig.\ref{fig1}). For instance, if the simulation approach 
to be employed is accurate first-principles methods the calculations are likely to be performed at 
zero temperature by considering perfectly ordered atomic structures. Such simulation conditions 
obviously differ from the actual experimental conditions. On the other hand, to simulate EC effects
directly at finite temperatures for realistic systems containing crystalline defects and/or other 
inhomogeneities one should use classical interatomic potentials or effective Hamiltonians, which may 
suffer from transferability issues and in general have modest predictive power. Fortunately, there 
are well-established simulation approaches that allow to achieve a suitable balance between 
computational accuracy and model reliability, and it is responsibility of the modeler to use them 
adequately for obtaining meaningful results. 

In this Chapter, we review computational techniques based on first-principles methods that can be 
used to estimate theoretically and predict EC effects. We start by briefly describing genuine 
first-principles methods, like density functional theory, and other practical approaches 
that rely on them to mimic polar materials, namely, bond-valence interatomic potentials and effective 
Hamiltonians. Next, the simulation strategies that are employed to assess EC effects with such 
methods, either directly or indirectly, are reviewed. We finalize the Chapter by providing some 
representative examples in which first-principles based methods have been used to discover and 
characterize original EC effects for multiferroic thin films [\onlinecite{cazorla18}], 
bulk hybrid organic-inorganic perovskites [\onlinecite{liu16}], and relaxor ferroelectrics 
[\onlinecite{jiang17}].

\section{Computational Methods}
\label{sec:methods}
The foundations of Density Functional Theory (DFT), a first-principles approach widely used in 
condensed-matter and materials science, bond-valence and classical interatomic potentials and 
effective Hamiltonians are briefly reviewed next. The two latter computational approaches are 
approximate and rely on the outputs of accurate first-principles methods (also called \emph{ab 
initio}), typically DFT. The computational cost of DFT methods is several orders of magnitude 
higher than those of interatomic potentials and effective Hamiltonians (Fig.\ref{fig1}), hence 
most of the times first-principles methods cannot be used straightforwardly to estimate EC effects.        

\subsection{$Ab$ $initio$ methods}
\label{subsec:abinitio}
In solids, the dynamics of the electrons and nuclei can be decoupled to a good approximation because 
their respective masses differ by several orders of magnitude. The wave function of the corresponding 
many-electron system, $\Psi ({\bf r}_{1}, {\bf r}_{2},...,{\bf r}_{N})$, therefore can be determined 
by solving the Schr\"{o}dinger equation corresponding to the non-relativistic Born-Oppenheimer Hamiltonian:
\begin{eqnarray}
H = -\frac{1}{2} \sum_{i} {\bf \nabla}^{2}_{i} - \sum_{I} \sum_{i} \frac{Z_{I}}{|{\bf R}_{I} - {\bf r}_{i}|} \nonumber \\
        + \frac{1}{2} \sum_{i} \sum_{j \neq i} \frac{1}{|{\bf r}_{i} - {\bf r}_{j}|}~,
\label{eq:BO-hamilton}
\end{eqnarray}
where $Z_{I}$ are the nuclear charges, ${\bf r}_{i}$ the positions of the electrons, and ${\bf R}_{I}$ 
the positions of the nuclei, which are considered fixed. In real materials, $\Psi$ is a complex mathematical 
function that in most cases is unknown. At the heart of any first-principles method is to find a good 
approximation for $\Psi$, or an equivalent quantity (e.g., the electronic density), that is manageable 
enough to perform calculations. Examples of \emph{ab initio} methods include density functional theory 
(DFT), M\o ller-Plesset perturbation theory (MP2), the coupled-cluster method with single, double and 
perturbative triple excitations [CCSD(T)], and quantum Monte Carlo (QMC), to cite just a few. Among these 
techniques, DFT methods are frequently applied to the study of ferroelectrics and multiferroics and for 
this reason we summarise their foundations in what follows.   

In 1965, Kohn and Sham developed a pioneering theory to effectively calculate the energy and properties 
of many-electron systems without the need of explicitly knowing $\Psi$ [\onlinecite{kohn65,sham66}]. 
The main idea underlying this theory, called density functional theory (DFT), is that the exact 
ground-state energy, $E$, and electron density, $n({\bf r})$, can be determined by solving an effective 
one-electron Schr\"odinger equation of the form:
\begin{equation}
H_{\rm KS} \Psi_{i \sigma} = \epsilon_{i \sigma} \Psi_{i \sigma}~, 
\label{eq:onelectron}
\end{equation}
where $H_{\rm KS}$ is the Kohn-Sham Hamiltonian, index $i$ labels different one-electron orbitals and 
$\sigma$ different spin states. The KS Hamiltonian can be expressed as:
\begin{equation}
H_{\rm KS} = -\frac{1}{2}\nabla^{2} + V_{ext}({\bf r}) + \int \frac{n({\bf r'})}{|{\bf r} - {\bf r'}|} d{\bf r'} + V_{xc}({\bf r})~,
\label{eq:heff}
\end{equation}
where
\begin{equation}
n({\bf r}) = \sum_{i \sigma} |\Psi_{i \sigma} ({\bf r})|^{2}~,
\label{eq:density}
\end{equation}
$V_{ext}$ represents an external field and $V_{xc} ({\bf r}) = \delta E_{xc} / \delta n ({\bf r})$
is the exchange-correlation potential.

The exchange-correlation energy has a purely quantum mechanical origin and can be defined as the
interaction energy difference between a quantum many-electron system and its classical counterpart.
Despite $E_{xc}$ represents a relatively small fraction of the total energy, this contribution is
extremely crucial for all materials and molecules because it acts directly on the bonding between
atoms. In general, $E_{xc}[n]$ is unknown and needs to be approximated. This is the only source
of fundamental error in DFT methods. In standard DFT approaches $E_{xc}[n]$ is approximated with 
the expression:
\begin{equation}
E_{xc}^{\rm approx}[n] = \int \epsilon_{xc}^{\rm approx}({\bf r}) n({\bf r}) d{\bf r}~,
\label{eq:excapprox}
\end{equation}
where $\epsilon_{xc}^{\rm approx}$ is made to depend on $n({\bf r})$, $\nabla n({\bf r})$,
and/or the electronic kinetic energy $\tau ({\bf r}) = \frac{1}{2} \sum_{i \sigma} 
|\nabla \Psi_{i \sigma} ({\bf r})|^{2}$~. Next, we summarise the basic aspects of the 
most popular $E_{xc} [n]$ functionals employed for computational analysis of archetypal 
ferroelectric and multiferroic materials. 

In local approaches (e.g., local density approximation --LDA--), $E_{xc}^{\rm approx}$ in 
Eq.(\ref{eq:excapprox}) is calculated by considering the exchange-correlation energy of an 
uniform electron gas with density $n({\bf r})$, $\epsilon_{xc}^{unif}$, which is known exactly 
from quantum Monte Carlo calculations [\onlinecite{ceperley80,perdew81}]. In order to deal 
with the non-uniformity of real electronic systems, the space is partitioned into infinitesimal 
volume elements that are considered to be locally uniform. In semi-local approaches (e.g., 
generalized gradient approximation --GGA--), $E_{xc}$ is approximated similarly to local approaches 
but $\epsilon_{xc}^{\rm approx}$ is made to depend also on the gradient of $n({\bf r})$  
[\onlinecite{perdew92,perdew96}]. Both local and semi-local approximations satisfy some exact 
$E_{xc}$ constraints and can work notably well for systems in which the electronic density 
varies slowly over the space (e.g., crystals). An extension of the GGA approach is provided 
by meta-GGA functionals, in which the non-interacting kinetic energy density is considered 
also as an energy functional input. An example of this latter type of functionals is the 
recently proposed meta-GGA SCAN [\onlinecite{scan,zhang18}].

Hybrid functionals comprise a combination of non-local exact Hartree-Fock and local exchange energies, 
together with semi-local correlation energies. The proportion in which both non-local and local exchange 
densities are mixed generally relies on empirical rules. The popular B3LYP approximation [\onlinecite{becke93}], 
for instance, takes a $20$\% of the exact HF exchange energy and the rest from the GGA and LDA functionals. 
Other well-known hybrid functionals are PBE0 [\onlinecite{adamo99}] and the range-separated HSE06 proposed 
by Scuseria and collaborators [\onlinecite{hse06}]. In contrast to local and semi-local functionals, 
hybrids describe the delocalisation of the exchange-correlation hole around an electron to some extent
so that they partially correct for electronic self-interaction errors (which are ubiquitous in standard
DFT) [\onlinecite{franchini14}]. This technical feature is specially useful to treat strongly correlated 
systems containing $d$ and $f$ electronic orbitals (e.g., transition-metal oxide perovskites) [\onlinecite{bilc08,evarestov12,cazorla16b,cazorla17c}].

\subsection{Bond-valence and classical interatomic potentials}
\label{subsec:potentials}
Using first-principles methods to describe the interactions between electrons and ions in crystals 
requires dedicated computational resources. In some cases, the interatomic interactions can be 
approximated satisfactorily by analytical functions known as classical interatomic potentials or 
force fields and consequently the simulations can be accelerated dramatically with respect to \emph{ab initio} 
calculations. Classical interaction models contain a number of parameters that are adjusted to 
reproduce experimental or \emph{ab initio} data, and their analytical expressions are constructed 
based on physical knowledge and intuition. The force matching method proposed by Ercolessi and Adams 
[\onlinecite{ercolessi94}] is an example of a classical potential fitting technique that is widely 
employed in the fields of condensed matter physics and materials science 
[\onlinecite{cazorla15c,mattoni15,hata17,cazorla18b}]. Nonetheless, the ways in which classical 
interatomic potentials are constructed are neither straightforward nor uniquely defined and the
thermodynamic intervals over which they remain reliable are limited.

A pairwise interaction model that has been employed to successfully simulate polar materials like 
BaTiO$_{3}$, LiNbO$_{3}$ and KNbO$_{3}$ at finite temperatures is the Coulomb--Buckingham (CB) potential, 
which adopts the simple form [\onlinecite{tinte99,jackson05,sepliarsky01,hashimoto14}]:
\begin{equation}
V_{\alpha \beta}(r_{ij}) = A_{\alpha \beta}e^{-\frac{r_{ij}}{\rho_{\alpha \beta}}} - \frac{C_{\alpha \beta}}{r_{ij}^{6}} + \frac{Z_{\alpha}Z_{\beta}}{r_{ij}}~,
\label{eq:BMH}
\end{equation}
where subscripts $\alpha$ and $\beta$ represent atomic species in the system, $r_{ij}$ the radial 
distance between a pair of $\alpha$ and $\beta$ atoms labelled $i$ and $j$ respectively, $Z$ ionic 
charges, and $A$, $\rho$ and $C$ are potential parameters. The CB potential is composed of three 
different energy contributions. The exponential term accounts for short-range repulsive forces 
resulting from the interactions between nearby electrons; the second term represents long-range 
attractive interactions arising from dispersive van der Waals forces; the third term is the usual 
Coulomb interaction between point charges. In order to describe atomic polarizability effects, 
``core-shell'' modelling can be done on top of the CB potential. In standard core-shell approaches, 
each atom is decomposed into a charged core, which interact with others through $V_{ij}$'s analogous 
to the expression shown in Eq.(\ref{eq:BMH}), and a charged shell that is bound harmonically to the 
core [\onlinecite{jackson05,sepliarsky01,hashimoto14,gambuzzi14}].  

Polar materials are characterized by a delicate balance between short-range and long-range forces, which
are respectively originated by complex transition metal (TM) $d$ and oxygen (O) $p$ electronic orbital 
hybridizations and Coulomb interactions between permanent electric dipoles and higher order moments 
[\onlinecite{cohen92}]. The simplicity of the pairwise interaction model enclosed in Eq.(\ref{eq:BMH}) 
may not be adequate to fully grasp the complexity of the TM--O bonding, which turns out to be critical to 
describe ferroelectricity and other relevant functional properties correctly. Bond-valence (BV) potentials 
represent an improvement with respect to the CB model because they can mimic chemical bonding in oxide 
perovskites and other complex materials more precisely [\onlinecite{haomin17}].   

A general BV potential for oxide perovskites is [\onlinecite{grinberg02,grinberg04,shin05,liu13}]:
\begin{equation}
V_{\rm BV}(r,\theta) = V_{\rm bind}(r) + V_{\rm charge}(r) + V_{\rm rep}(r) + V_{\rm nl}(\theta)~,
\label{eq:BV}
\end{equation}
where the first term in the right-hand side represents the bond-valence potential energy, the second 
the Coulomb potential energy, the third the repulsive potential energy, and the fourth an angle potential 
energy (to prevent unphysically large distortions of the oxygen octahedra). The bond-valence energy 
term generally is expressed as:
\begin{equation}
V_{\rm bind}(r) = \sum_{\alpha = 1}^{N_{s}} S_{\alpha} \sum_{i = 1}^{N_{\alpha}} \mid V_{i \alpha}(r_{i}) - V_{\alpha} \mid^{\gamma_{\alpha}}, 
\label{eq:BV-binding}
\end{equation}
with
\begin{equation}
V_{i \alpha}(r_{i}) = \sum_{\beta = 1}^{N_{s}} \sum_{j}^{NN} \left( \frac{r_{0}^{\alpha \beta}}{r_{ij}^{\alpha \beta}} \right)^{C_{\alpha \beta}},
\label{eq:BV-binding-II}
\end{equation}
where $N_{s}$ represents the number of atomic species in the system (e.g., $3$ for PbTiO$_{3}$), $S_{\alpha}$ 
are fitting parameters, $N_{\alpha}$ the number of $\alpha$ atoms, $V_{\alpha}$ the desired atomic valence for 
ion $\alpha$, $\gamma_{\alpha}$ fitting parameters typically set to $1$, $j$ an atomic index that runs over 
nearest-neighbour (NN) ions, $r_{0}^{\alpha \beta}$ and $C_{\alpha \beta}$ parameters determined by empirical 
rules, and $r_{ij}^{\alpha \beta}$ the radial distance between ions $i$ and $j$. 

For the repulsive energy term, $V_{\rm rep}$, the following expression normally is employed:
\begin{equation}
V_{\rm rep}(r) = \epsilon \sum_{\alpha = 1}^{N_{s}} \sum_{i = 1}^{N_{\alpha}} \sum_{\beta = 1}^{N_{s}} \sum_{j = 1}^{N_{\beta}} \left( \frac{B_{\alpha \beta}}{r_{ij}^{\alpha \beta}} \right)^{12},
\label{eq:BV-rp}
\end{equation}    
where $\epsilon$ and $B_{\alpha \beta}$ are fitting parameters. Meanwhile, an harmonic function is used
for the angle potential energy that reads:
\begin{equation}
V_{\rm nl}(\theta) = k \sum_{i = 1}^{N_{oct}} \left( \theta_{i,x}^{2} + \theta_{i,y}^{2} + \theta_{i,z}^{2} 
                                              \right), 
\label{eq:BV-angle}
\end{equation}
where $k$ is a fitting parameter, $N_{oct}$ the number of the oxygen octahedra, and $\lbrace \theta_{i,\gamma} 
\rbrace$ the angles between the oxygen octahedral axes and the system reference axes.  

Reliable BV potentials have been developed for archetypal ferroelectric and piezoelectric materials 
like BaTiO$_{3}$, PbTiO$_{3}$, and PbZr$_{0.2}$Ti$_{0.8}$O$_{3}$ 
[\onlinecite{grinberg02,grinberg04,shin05,liu13,xu15}]. Recently, it has been predicted based on the 
outcomes of molecular dynamics simulations performed with BV potentials that ultrafast electric-field 
pulses can induce giant and inverse EC effects (i.e., $\Delta T < -10$~K) in bulk BaTiO$_{3}$ and PbTiO$_{3}$ 
[\onlinecite{qi18}]. Such caloric effects occur in the time scale of few picoseconds hence they could 
be exploited for the design of fast solid-state cooling processes.

\subsection{Effective Hamiltonians}
\label{subsec:Heff}
Another first-principles based approach that has proven very successful in describing ferroelectric
oxide perovskites is the effective Hamiltonian method [\onlinecite{zhong95,paul17,prosandeev15}].
In this approach, a subset of degrees of freedom that is relevant to the observed phase transitions
is first selected. The parameters defining the effective Hamiltonian are then determined by performing 
accurate zero-temperature first-principles calculations. Finally, finite-$T$ simulations are undertaken 
to assess displacive-like phase transition governed by the constructed effective Hamiltonian.  
In the effective Hamiltonian approach, the energy surface of the polar crystal is approximated by
a low-order Taylor expansion of the energy surface of a high-symmetry non-polar cubic phase, which
is observed in some archetypal ferroelectrics at high temperatures. 

A typical effective Hamiltonian consists of five energy contributions: a local-mode self-energy, 
a long-range dipole-dipole interaction, a short-range interaction between soft modes, an elastic energy, 
and an interaction between the local modes and local strain [\onlinecite{zhong95}]. Analytically, this 
model is expressed as:
\begin{eqnarray}
E^{\rm tot} & = & E^{\rm self}(\lbrace{{\bf u}\rbrace}) + E^{\rm dipol}(\lbrace{{\bf u}\rbrace}) + E^{\rm short}(\lbrace{{\bf u}\rbrace}) + \nonumber \\
            &   & E^{\rm elast}(\lbrace{ \eta_{l} \rbrace}) + E^{\rm int}(\lbrace{ \eta_{l} \rbrace}, \lbrace{{\bf u}\rbrace})~,
\label{eq:effecHam}
\end{eqnarray}  
where $\lbrace{{\bf u}\rbrace}$ are the amplitude vector of the lowest-energy transversal optical 
$\Gamma$--phonon modes of the parent cubic phase, and $\lbrace{ \eta_{l} \rbrace}$ the corresponding 
six-component local strain tensor. 

The local-mode self-energy term adopts the form:
\begin{eqnarray}
	E^{\rm self}(\lbrace{{\bf u}\rbrace}) & = & \sum_{i} \kappa u_{i}^{2} +  \alpha u_{i}^{4} + \nonumber \\
	& & \gamma \left ( u_{ix}^{2}u_{iy}^{2} + u_{ix}^{2}u_{iz}^{2} + u_{iy}^{2}u_{iz}^{2} \right)~, 
\label{eq:self-energy}
\end{eqnarray}
where $u_{i} = |{\bf u}_{i}|$ and $\kappa$, $\alpha$ and $\gamma$ are expansion parameters to be determined 
from first-principles calculations. For the long-range energy term, only dipole-dipole interactions are 
considered and the following expression is employed:
\begin{equation}
	E^{\rm dipol}(\lbrace{{\bf u}\rbrace}) = \frac{{\cal Z}^{2}}{\epsilon_{\infty}} \sum_{i < j} 
	\frac{ {\bf u}_{i}{\bf u}_{j} - 3 \left({\bf \hat{r}}_{ij} {\bf u}_{i}\right)\left({\bf \hat{r}}_{ij} {\bf u}_{j}\right)}{r_{ij}^3}~, 
\label{eq:dipole}
\end{equation}
where ${\cal Z}$ is the Born effective charge associated with the soft mode, $\epsilon_{\infty}$ the optical
dielectric constant of the material, ${\bf r}_{ij} \equiv {\bf r}_{i} - {\bf r}_{j}$ the distance
vector between different unit cells, and ${\bf \hat{r}}_{ij} \equiv {\bf r}_{ij} / r_{ij}$. 
To express the short-range interactions between neighboring local modes, a formula that is reminiscent 
of the spin Heisenberg model is adopted:
\begin{equation}
E^{\rm short}(\lbrace{{\bf u}\rbrace}) = \frac{1}{2} \sum_{i \neq j} \sum_{\alpha \beta} J_{ij,\alpha \beta} u_{i \alpha} u_{j \beta}~, 
\label{eq:short}
\end{equation}	
which typically applies to first, second and third neighbouring unit cells, and where the coupling matrix 
$J_{ij,\alpha \beta}$ depends on $r_{ij}$ and decays rapidly with increasing distance. Meanwhile, the 
elastic energy and elastic-mode interaction energy terms are deduced by considering symmetry and 
stress--strain relationships of the parent cubic phase (e.g., elastic constants). For instance, 
$E^{\rm elast}(\lbrace{ \eta_{l} \rbrace})$ is expressed as a sum of homogeneous and inhomogenous 
deformations that allow to change the shape and volume of the simulation cell, while the elastic-mode 
interaction energy adopts the form:
\begin{eqnarray}
E^{\rm int}(\lbrace{ \eta_{l} \rbrace}, \lbrace{{\bf u}\rbrace})& = & \nonumber \frac{1}{2} \sum_{i} \sum_{l \alpha \beta} B_{l \alpha \beta} \times \\ 
& & \eta_{l}\left({\bf r}_{i}\right) {u}_{\alpha}\left({\bf r}_{i}\right) {u}_{\beta}\left({\bf r}_{i}\right)~, 
\label{eq:elasticint}
\end{eqnarray} 
where $B_{l \alpha \beta}$ are strain--mode coupling constants.  

To put some numbers on the construction of the $E^{\rm tot}$ functional expressed in Eq.(\ref{eq:effecHam}), 
a total of $18$ expansion parameters estimated with first-principles methods are required to obtain a minimum
reliable effective Hamiltonian model for BaTiO$_{3}$ [\onlinecite{zhong95}]. Depending on the material and 
physical phenomena to be simulated more complexity can be added to the $E^{\rm tot}$ expression, although the 
number of involved expansion parameters can increase very rapidly [\onlinecite{paul17,prosandeev15}]. A version 
of the effective Hamiltonian approach for oxide perovskite has been already implemented in the freely available 
code package FeRam [\onlinecite{feram}]. 

Effective Hamiltonians, as any other method, present some limitations. First, because they focus 
on certain degrees of freedom they may not be accurate enough to capture the full range of behavior 
of the material under study, especially for configurations beyond the model training database. Second, 
an appropriate selection of the degrees of freedom requires significant insight into the material, 
which may retard the development of such models for new systems. Third, first-order phase transitions 
characterised by large symmetry and volume changes cannot be simulated with effective Hamiltonians 
(recall the fundamental approximation of a low-order Taylor expansion around a presumed reference 
cubic phase). And fourth, the connectivity between atoms needs to be fixed for all the simulated phases 
thus variations in the atomic environment cannot be reproduced. 
 
Recently, an extension of the effective Hamiltonian method has been developed, the so-called 
``second-principles'' approach, which may overcome some of the technical issues just mentioned 
[\onlinecite{scaleup,zubko16}]. In the second-principles approach, the parameters of the effective
Hamiltonian model are computed in a very fast and efficient way by recasting the $E^{\rm tot}$ fit 
to a training set of first-principles data into a simple matrix diagonalization problem. Specifically, 
the interactions that are most relevant to reproduce the first-principles training-set data are 
selected automatically from a pool that virtually includes all possible coupling terms. The 
second-principles method has been already implemented in the freely available code package SCALE-UP 
[\onlinecite{scaleup}].

\section{Indirect estimation of the EC effect}
\label{sec:indirect}
The isothermal entropy change that a polar material undergoes under the action of a varying external
electric field can be expressed by means of the Maxwell relations as [\onlinecite{moya14}]:
\begin{equation}
\Delta S (T, {\cal E}_{f}) = \int_{0}^{{\cal E}_{f}} \left( \frac{dP}{dT}\right)_{{\cal E}} d{\cal E}~,
\label{eq:deltaS}
\end{equation}
where $P$ represents the electric polarization of the system and ${\cal E}$ the applied electric field. 
Likewise, the corresponding adiabatic temperature shift can be estimated as:
\begin{eqnarray}
\Delta T (T, {\cal E}_{f}) & = & -\int_{0}^{{\cal E}_{f}} \frac{T}{C_{\cal E}(T)} dS \nonumber \\ 
                           & \approx & -\frac{T}{C_{0}(T)} \Delta S(T, {\cal E}_{f})~,
\label{eq:deltaT}
\end{eqnarray}
where $C_{\cal E}(T) \equiv \left( \frac{dU}{dT} \right)_{\cal E}$ is the heat capacity of the 
system at fixed electric field ${\cal E}$ and $U$ the internal energy. Typically, $\Delta S$ and 
$\Delta T$ are large in the vicinity of a ${\cal E}$--induced phase transition because at such 
conditions the $T$--induced variation of $P$ is most significant. 

Equations (\ref{eq:deltaS}) and (\ref{eq:deltaT}) show that the isothermal entropy and adiabatic 
temperature shifts associated with the electrocaloric (EC) effect can be estimated indirectly by knowing 
the heat capacity of the system and the pyroelectric coefficient, $\alpha_{\cal E} \equiv \left( \frac{dP}{dT} 
\right)_{\cal E}$. As we explain next, indirect estimation of EC effects can be achieved theoretically 
by using (i)~classical interaction potentials and effective Hamiltonian models in classical molecular dynamics 
and Monte Carlo simulations, and (ii)~first-principles methods in combination with quasi-harmonic approaches 
(Sec.\ref{sec:methods} and Fig.\ref{fig1}). It is worth noting that indirect estimation of EC effects 
is only meaningful in the context of second-order phase transitions since otherwise the pyroelectric 
coefficient $\alpha_{\cal E}$ is ill defined at the transition point and consequently Eq.(\ref{eq:deltaS}) 
cannot be estimated numerically (see next section for the estimation of EC effects associated with 
first-order phase transitions).  

In both molecular dynamics (MD) and Monte Carlo (MC) simulations it is possible to simulate 
the system of interest at the desired $T$, ${\cal E}$ and pressure (or volume) conditions. MD
simulations are deterministic and rely on the numerical discretization and integration of Newton's second 
law of motion [\onlinecite{frenkel01}]. MC simulations, by contrast, are stochastic and rely on efficient 
sampling of the Maxwell-Boltzmann probability distribution [\onlinecite{metropolis53}]. Nevertheless, 
when the technical parameters in MD and MC simulations are selected to ensure proper convergence
the average results obtained in both types of simulations should be equivalent. Typically, 
large system sizes ($N \sim 10^{3}$--$10^{7}$) and long simulation times ($\tau \sim 1$--$10$~ns) 
are required to obtain well converged results in MD and MC simulations. As a consequence, 
first-principles methods normally are excluded from such type of finite-$T$ approaches and 
 bond valence/classical interaction potentials and effective Hamiltonians remain as the 
\emph{ab initio} based alternatives for modelling the interatomic interactions 
[\onlinecite{liu16,jiang17,marathe16,jiang18}].     

The heat capacity and pyroelectric coefficient of a polar crystal can be estimated efficiently 
with both MD and MC methods by computing average internal energies and atomic positions (or 
local modes) in simulations performed at different temperatures. Alternatively, $\Delta S$
can be estimated in a quasi-direct fashion by integrating the specific heat obtained at constant 
electric field like: 
\begin{equation}
\Delta S (T, {\cal E}_{f}) = \int_{T_{0}}^{T} \frac{C_{0}(T') - C_{{\cal E}_{f}}(T')}{T'}~dT'~,
\label{eq:deltaS-quasi}
\end{equation} 
where $T_{0}$ needs to be sufficiently low in practice so that the condition $S (T_{0}, 0) \approx 
S (T_{0}, {\cal E}_{f})$ is fulfilled [\onlinecite{marathe16}]. In the context of effective 
Hamiltonians, the following expression also can be used to estimate adiabatic temperature shifts 
produced by electric fields [\onlinecite{jiang17}]: 
\begin{equation}
\Delta T (T, {\cal E}_{f}) = -\gamma_{T} \int_{0}^{{\cal E}_{f}} 
                           \frac{\langle u \cdot U \rangle - \langle u \rangle \cdot \langle U \rangle}{\langle U^{2} \rangle - \langle U \rangle^{2}}~d{\cal E}~, 
\label{eq:deltaT-effec}
\end{equation} 
where $\gamma_{T} \equiv {\cal Z} a_{0} N T$, ${\cal Z}$ is the Born effective charge associated with the soft mode, $a_{0}$ the lattice parameter
of the cubic unit cell, $N$ the number of sites in the supercell, $u$ the module of the average supercell
local mode, $U$ the total energy, and $\langle \cdots \rangle$ denotes thermal average.

Despite the high computational cost associated with first-principles methods, it is also possible to 
estimate EC effects indirectly with them by means of the quasi-harmonic approximation (QHA). 
In the quasi-harmonic approach, the internal energy of a crystal is expressed as [\onlinecite{harmonic1,harmonic2,harmonic3}]:
\begin{equation}
E_{\rm harm}(T) = E_{0} + \frac{1}{2} \sum_{mn} \Xi_{mn} u_{m}(T) u_{n}(T)~,
\label{eq:harmonicE}
\end{equation}
where $E_{0}$ corresponds to the static energy, $\Xi_{mn}$ the force-constant matrix and $\lbrace u 
\rbrace$ atomic displacements that depend on $T$. The phonon frequencies of the crystal, $\lbrace
\omega_{{\bf q}s} \rbrace$, and corresponding normal mode amplitudes, $\lbrace Q_{{\bf q}s} \rbrace$, 
are calculated from the diagonalization of the dynamical matrix obtained from $\Xi_{mn}$.

According to well-established theories [\onlinecite{lang05}], the pyroelectric coefficient of a crystal 
subject to constant stress, $\sigma$, can be decomposed into two contributions: the primary contribution 
obtained at constant strain, $\alpha_{{\cal E}(\eta)}$, and the secondary contribution associated with 
the material thermal expansion, $\alpha_{{\cal E}(T)}$. These two parts can be expressed analytically as 
[\onlinecite{liu18,liu17}]:   
\begin{eqnarray}
\alpha_{{\cal E}} = \alpha_{{\cal E}(\eta)} + \alpha_{{\cal E}(T)} & = & \left( \frac{\partial P}{\partial T} \right)_{\eta} + \nonumber \\ 
& & \sum_{i} \left( \frac{\partial P}{\partial \eta_{i}} \right)_{T}  \left( \frac{d \eta_{i}}{d T} \right)_{\sigma}~.
\label{eq:pyroelectric}
\end{eqnarray}
The secondary contribution to $\alpha_{{\cal E}}$ corresponds to the pyroelectricity induced by the thermal
expansion, which can be obtained from the material piezoelectric stress constants and volume expansion 
coefficients computed with first-principles methods. The primary contribution to $\alpha_{{\cal E}}$ 
corresponds to the ``clamped-lattice'' pyroelectricity, which results from holding the lattice parameters 
of the crystal fixed. 

The electric polarization can be expanded in terms of the vibrational normal mode amplitudes, 
$\lbrace Q_{{\bf q}s} \rbrace$, according to the formula [\onlinecite{liu18,liu17,born45}]:
\begin{equation}
P (T) = P (0) + \sum_{{\bf q}s}\frac{\partial P}{\partial Q_{{\bf q}s}} \langle Q_{{\bf q}s} \rangle~, 
\label{eq:expansionP}
\end{equation}
where the first term represents the electric polarization in the absence of thermal fluctuations, 
which can be readily calculated with first-principles methods [\onlinecite{cazorla15}]. The primary 
contribution to the pyroelectric coefficient in Eq.(\ref{eq:pyroelectric}) then can be expressed 
as [\onlinecite{liu18,liu17,born45}]:
\begin{equation}
\alpha_{{\cal E}(\eta)}  = 
       \sum_{s} \frac{\partial P}{\partial Q_{\Gamma s}} \frac{d\langle Q_{\Gamma s} \rangle}{dT} + 
       \sum_{{\bf q}s} \frac{\partial^{2} P}{\partial Q_{{\bf q} s}^{2}} \frac{d\langle Q_{{\bf q} s}^{2} \rangle}{dT}~,
\label{eq:primary}
\end{equation}
where only polar $\Gamma$-optical survive in the first term. By using the QHA and first-principles 
simulations, one can estimate the temperature derivatives in Eq.(\ref{eq:primary}) via the expressions 
[\onlinecite{liu18,liu17}]:
\begin{eqnarray}
\langle Q_{\Gamma s} \rangle & = & \sum_{{\bf q}l} \frac{\hbar}{2} \frac{\left( 2n_{{\bf q}l} + 1 \right)}{\omega_{\Gamma s}^{2}} \frac{\partial \omega_{{\bf q}l}}{\partial Q_{\Gamma s}}   \\
\langle Q_{{\bf q} s}^{2} \rangle & = & \frac{\hbar}{2} \frac{\left( 2n_{{\bf q}s} + 1 \right)}{\omega_{{\bf q} s}}~, 
\label{eq:t-deriv-primary}
\end{eqnarray}
where $n$ is the Bose--Einstein distribution function, which depends explicitly on the temperature 
[\onlinecite{harmonic1}]. 
The pyroelectric properties of several semiconductor (e.g., GaN and ZnO [\onlinecite{jiang17}]) and 
two-dimensional materials (e.g., GeS and MoSSe [\onlinecite{jiang18}]) have been estimated 
successfully with the explained combination of \emph{ab initio} methods and the quasi-harmonic approach.  

Likewise, the heat capacity of a crystal can be estimated within the quasi-harmonic approximation by 
using the formula [\onlinecite{cazorla19b}]: 
\begin{equation}
C_{0}(T) = \frac{1}{N_{q}}~\sum_{{\bf q}s} \frac{\left( \hbar \omega_{{\bf q}s} \right)^{2}}{k_{B} T^{2}} \frac{e^{\frac{\hbar \omega_{{\bf q}s}}{k_{B} T}}}{\left( e^{\frac{\hbar \omega_{{\bf q}s}}{k_{B} T}} - 1 \right)^{2}} ~, 
\label{eq:heatc}
\end{equation}
where $N_{q}$ is the total number of wave vectors used for integration in the first Brillouin zone 
and the summation runs over all wave vectors ${\bf q}$ and phonon branches $s$. Thus, by using 
Eqs.(\ref{eq:pyroelectric})--(\ref{eq:heatc}), that is, by determining $\alpha_{{\cal E}}$ and $C_{0}$ 
and their dependence on temperature, in principle it is possible to estimate EC effects indirectly with 
first-principles methods and remarkable accuracy without the need to perform finite-$T$ MD or MC simulations.

\section{Clausius-Clapeyron method}
\label{sec:clausius}
In the case that the ${\cal E}$--field induced phase transition of interest presents a marked
first-order character, the corresponding isothermal entropy change can be estimated with the 
Clausius-Clapeyron method as [\onlinecite{moya14}]:
\begin{equation}
\Delta S(T) = -\Delta P(T) \frac{d{\cal E}_{c}}{dT}~, 
\label{eq:clausiusS}
\end{equation}
where $\Delta P(T)$ is the change in polarization along the electric field direction, and 
${\cal E}_{c}(T)$ the critical electric field inducing the phase transformation.  

In the presence of an electric field and by assuming zero-pressure conditions, the thermodynamic 
potential that appropriately describes the stability of a particular phase is the Gibbs free energy 
defined as $G = F - \bm{{\cal E}} \cdot \bm{P}$, where $F$ represents the Helmholtz free energy 
(where $F \equiv U - TS$ and $U$ is the internal energy). Accordingly, the thermodynamic condition 
that determines the ${\cal E}$--induced phase transition between states $A$ and $B$ is $G^{A}(T,{\cal E}_{c}) 
= G^{B}(T, {\cal E}_{c})$. The approximate value of the corresponding critical electric field then can be 
estimated as:
\begin{equation}
{\cal E}_{c}(T) \approx \frac{\Delta F(T)}{\Delta P(T)}~,
\label{eq:clausiusE}
\end{equation}
where $\Delta F$ represents the Helmholtz free energy difference between the two states, and $\Delta P$ 
their electric polarization difference along the electric field direction. By knowing $\Delta F(T)$ and 
$\Delta P(T)$, then one can calculate $\Delta S$ and, if the heat capacity of the system is also 
known, $\Delta T$ as a function of temperature. 

The Helmholtz free energy of the competing polymorphs in some cases can be estimated accurately as a 
function of temperature with first-principles methods and the quasi-harmonic approximation (QHA) 
[\onlinecite{harmonic1,harmonic2,harmonic3}]. In particular, the Helmoltz free energy associated 
with the lattice vibrations, $F_{\rm vib}$, is calculated by finding the phonon frequencies of the 
crystal, $\lbrace \omega_{{\bf q}s} \rbrace$, and subsequently using the formula:
\begin{equation}
F_{\rm vib} (T) = \frac{1}{N_{q}}~k_{B} T \sum_{{\bf q}s}\ln\left[ 2\sinh \left( \frac{\hbar \omega_{{\bf q}s}}{2k_{\rm B}T} \right) \right]~,
\label{eq:qha-vib}
\end{equation}
where $N_{q}$ is the total number of wave vectors used for integration in the first Brillouin zone, and 
the summation runs over all wave vectors ${\bf q}$ and phonon branches $s$. The total Helmholtz free 
energy of the system finally can be estimated as:
\begin{eqnarray}
F_{\rm harm} (T) = E_{0} + F_{\rm vib} (T)~,
\label{eq:helmholtz}
\end{eqnarray}
where $E_{0}$ is the static energy calculated with the atoms frozen on their equilibrium lattice positions.

Since the electric polarization difference between polymorphs $A$ and $B$ and their heat capacities also 
can be estimated as a function of temperature within the quasi-harmonic approximation (Sec.\ref{sec:indirect}), 
the adiabatic temperature and isothermal entropy changes in Eqs.(\ref{eq:deltaT}) and (\ref{eq:clausiusS})
in principle can be determined with first-principles methods. In fact, the described QHA method has been 
employed recently to predict large EC effects in multiferroic BiCoO$_{3}$ thin films caused by abrupt 
first-order phase transitions [\onlinecite{cazorla18}] (this illustrative case will explained in detail
in Sec.\ref{subsec:bicoo3}). 

Nevertheless, anharmonic effects, which mostly are not accounted for by quasi-harmonic approaches, can be 
very important in ferroelectric materials (e.g., BaTiO$_{3}$) and consequently the QHA may not be adequate 
to describe them correctly (for instance, when imaginary frequencies appear in the vibrational phonon 
spectra Eq.(\ref{eq:qha-vib}) cannot be used [\onlinecite{harmonic1}]). In this case, one still can employ 
genuine \emph{ab initio} methods to compute anharmonic free energies for the relevant polymorphs with 
methods like thermodynamic integration from reference models [\onlinecite{taioli07,cazorla07,cazorla12}] 
and self-consistent phonon approaches [\onlinecite{scp1,scp2,scp3}]. The computational load associated 
with these latter \emph{ab initio} anharmonic methods, however, is huge and to the best of our knowledge 
they have not been used for the simulation of EC effects to this day.

\section{Direct estimation of the EC effect}
\label{sec:direct}
MD and MC simulations allow to estimate EC effects directly, that is, by avoiding the use of the Maxwell 
relations. This modality of EC simulation may be beneficial for cases in which (i)~the definition of 
the electric polarization poses some ambiguity (e.g., disordered crystals [\onlinecite{cazorla15b}]),  
and/or (ii)~the polar degrees of freedom are strongly coupled with the lattice strain (e.g., good
piezoelectric materials) since in that case Eq.(\ref{eq:deltaS}) neglects possible secondary EC 
effects arising from ${\cal E}$-induced structural distortions [\onlinecite{wang20}]. On the 
other hand, this type of simulations only can be performed feasibly with classical interatomic/bond 
valence potentials and effective Hamiltonians because involve explicit simulation of $T \neq 0$ conditions, 
and apply exclusively to EC effects deriving from second-order phase transitions. 

\begin{figure}[t]
\centerline
        {\includegraphics[width=1.00\linewidth]{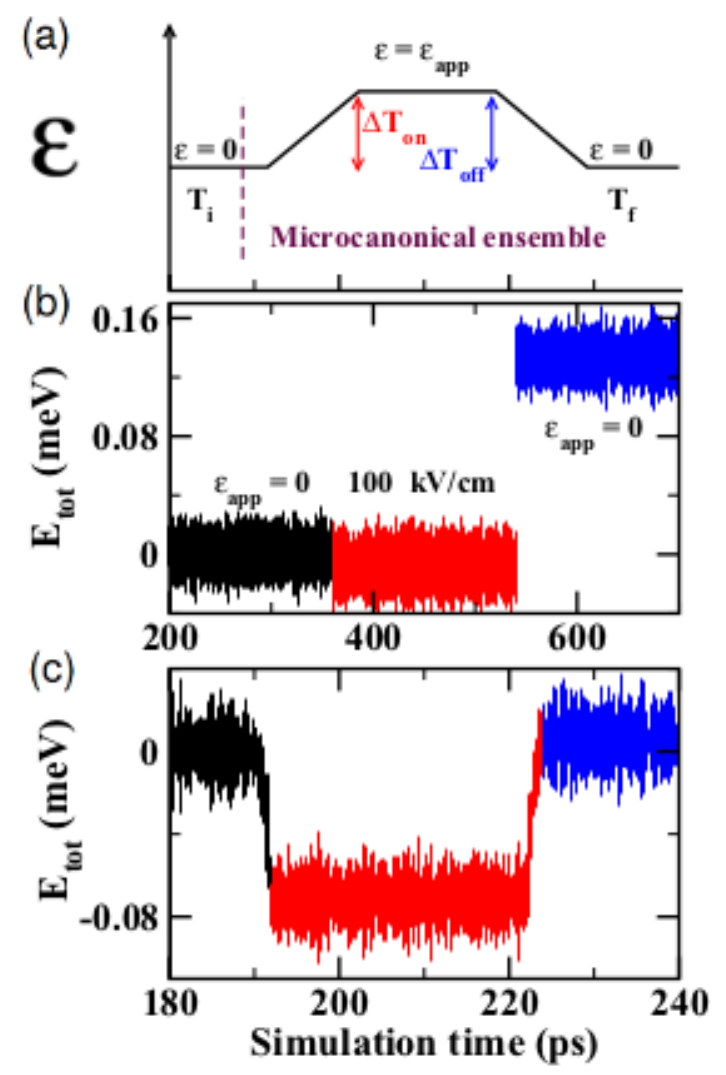}}
\caption{(a)~Sketch of the simulation ${\cal E}$-ramping protocol employed for direct estimation 
         of EC effects. Total energy of the system as a function of MD steps for (b)~instantaneous 
         field switching on/switching off and (c)~slow field ramping, respectively. The process
         simulated in (b) is irreversible and in (c) practically reversible. Reproduced from 
         work [\onlinecite{marathe16}].}
\label{fig2}
\end{figure} 

Direct estimation  of EC effects have been performed for a number of different systems including 
hybrid organic-inorganic perovskites [\onlinecite{liu16}] and multifunctional oxides 
[\onlinecite{qi18,jiang17,marathe16,jiang18}]. In this type of simulations, the electric field needs 
to be applied/removed very slowly so that the electrical polarization can follow adiabatically the 
external field. On the contrary, the simulated EC process is irreversible to some extent and the 
resulting adiabatic temperature shifts are not meaningful (Fig.\ref{fig2}). 

Initially, the system is thermalized at the temperature of interest, $T_{i}$, and zero electric field 
in the canonical, $(N,V,T)$, or the isobaric-isothermal, $(N,p,T)$, ensemble (where $p$ represents the 
hydrostatic pressure). After thermalization, the simulation is switched to the microcanonical or the 
isoenthalpic-isobaric ensemble [$(N,E,V)$ or $(N,H,p)$ where $E$ and $H$ represent the internal energy 
and enthalpy, respectively]. At this stage the electric field is ramped up to the desired value slowly 
enough to guarantee adiabaticity, and the accompanying temperature change, $\Delta T_{\rm on}$, is 
monitored. The system is simulated under these conditions for some time. Subsequently, the electric 
field is ramped down to zero slowly enough again to guarantee adiabaticity and the corresponding 
temperature change, $\Delta T_{\rm off}$, and final temperature, $T_{f}$, are monitored. 

Under the condition that the system remains in thermal equilibrium during the entire described 
simulation cycle, it will follow that $T_{i} = T_{f}$ and $\Delta T_{\rm on} = \Delta T_{\rm off}$ 
within their corresponding statistical uncertainties (Fig.\ref{fig2}c and [\onlinecite{marathe16}]). 
In such a case, either $\Delta T_{\rm on}$ or $\Delta T_{\rm off}$ can be identified with the adiabatic 
temperature change associated with the simulated EC process. 

\begin{figure*}[t]
\centerline
        {\includegraphics[width=0.90\linewidth]{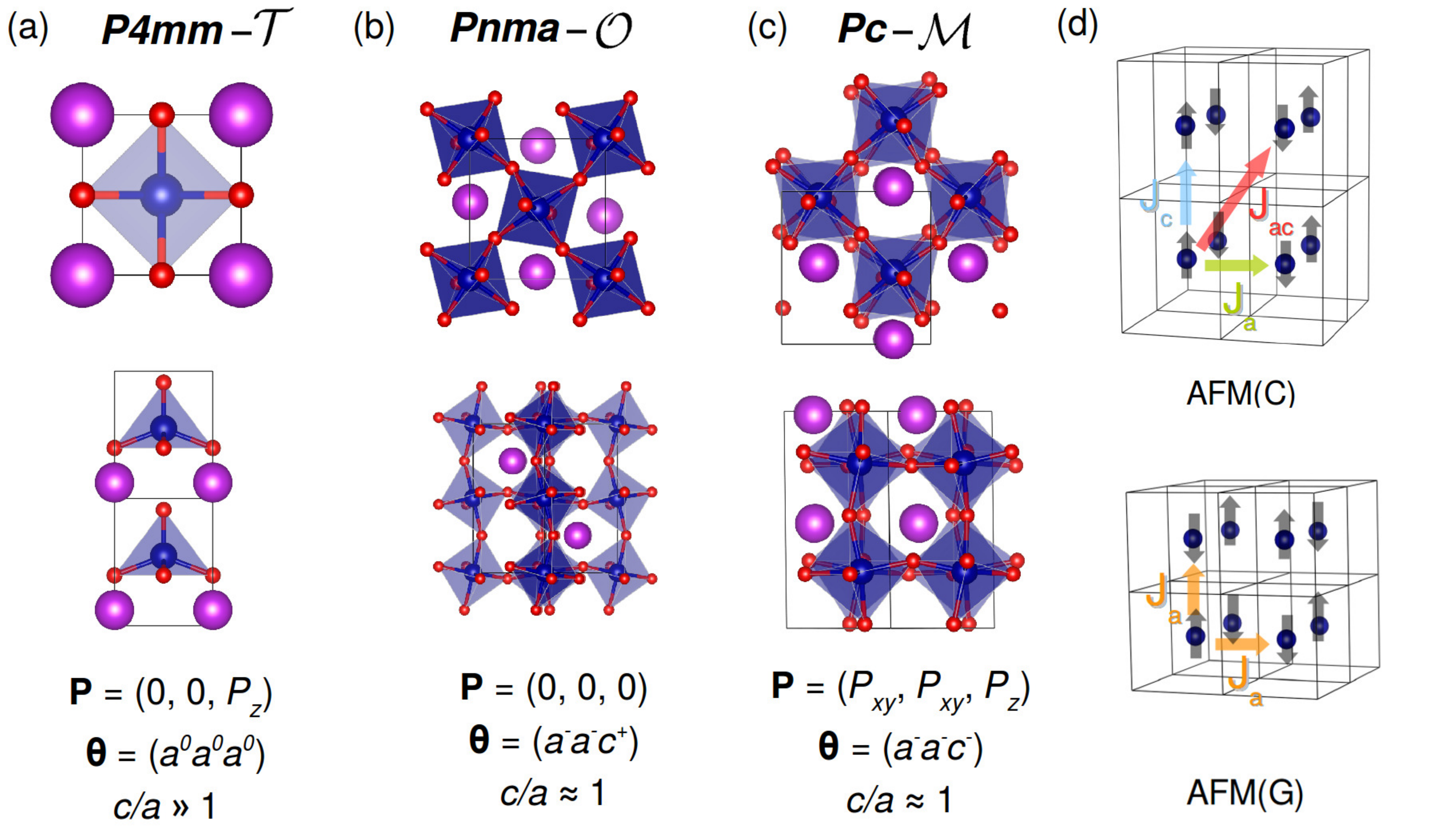}}
\caption{Structural, ferroelectric, and magnetic properties of energetically
         competitive bulk BCO polymorphs. (a)~Tetragonal $P4mm$ (${\cal T}$),
         (b)~orthorhombic $Pnma$ (${\cal O}$), and (c)~monoclinic $Pc$
         (${\cal M}$). (d)~Sketch of the spin configurations and exchange
         constants considered for the ${\cal T}$--AFM(C) and ${\cal O}$--AFM(G)
         Heisenberg spin models. Electrical polarizations {\bf P} are referred
         to pseudocubic Cartesian axis, and oxygen-octahedra rotation patterns
         $\boldsymbol{\theta}$ are expressed in Glazer's notation. Reproduced from
         work [\onlinecite{cazorla18}].}
\label{fig:bco-tf-0}

\end{figure*}
\section{Representative examples}
\label{sec:examples}
In this section, we describe three illustrative cases in which original EC effects have been simulated
and quantified with first-principles based methods. The systems in which those EC effects have been
predicted are technologically relevant, namely, multiferroic BiCoO$_{3}$ thin films (\emph{ab initio} 
methods in combination with the quasi-harmonic approximation), the organic-inorganic halide perovskite 
CH$_{3}$NH$_{3}$PbI$_{3}$ (classical force fields and molecular dynamics simulations) and the relaxor 
ferroelectric BaZr$_{1-x}$Ti$_{x}$O$_{3}$ (effective Hamiltonians and molecular dynamics and Monte Carlo 
simulations). Analogous EC simulation success can be achieved for other similar polar materials. 

\begin{figure*}[t]
\centerline
        {\includegraphics[width=0.90\linewidth]{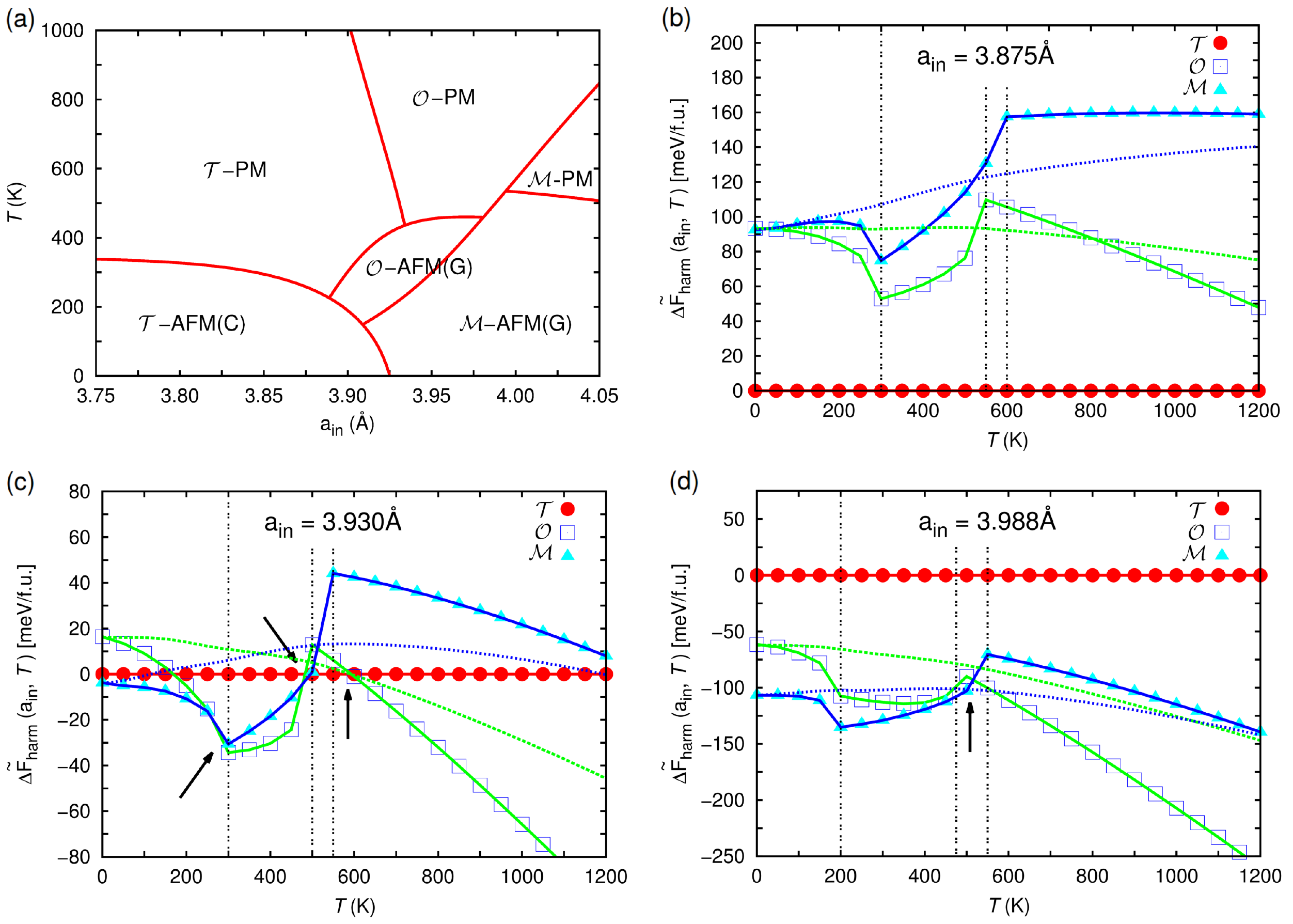}}
\caption{$T$--$a_{\rm in}$ phase diagram and free-energy differences in $(100)$-oriented BCO thin
  films. (a)~Phase diagram as a function of temperature and in-plane lattice parameter.
  (b)~Helmholtz free-energy differences among competitive polymorphs at $a_{\rm in} = 3.875$~\AA,
  (c)~$a_{\rm in} = 3.930$~\AA, (d)~$a_{\rm in} = 3.988$~\AA.
  Vertical lines and black arrows indicate $T$--induced magnetic and structural phase transitions,
  respectively. Blue and green dotted lines correspond to Helmholtz free-energy differences
  for the ${\cal M}$ and ${\cal O}$ phases, respectively, calculated without considering spin-phonon
  coupling effects. Reproduced from work [\onlinecite{cazorla18}].}
\label{fig:bco-tf-I}
\end{figure*}  
 
\subsection{Multiferroic BiCoO$_{3}$ thin films}
\label{subsec:bicoo3}
In multiferroic materials several order parameters coexist, typically ferroelectricity and magnetism, and 
are coupled to each other. Such unique properties convert multiferroics into promising materials for 
applications in memory devices, sensors and solid state cooling [\onlinecite{cazorla19b}], to cite few 
examples. The archetypal multiferroic compound is the oxide perovskite BiFeO$_{3}$, which presents 
a rhombohedral ground-state phase characterised by an electric polarisation of $\sim 60$~$\mu$C/cm$^{2}$ 
and antiferromagnetic (AFM) spin ordering of G-type [\onlinecite{cazorla13}] (Fig.\ref{fig:bco-tf-0}). 

Bulk BiCoO$_{3}$ (BCO) is another multiferroic material in which ferroelectricity and antiferromagnetism
coexist at ambient conditions. The stable phase of BCO is ferroelectric (FE) and tetragonal ${\cal T}$,
with a significantly large out-of-plane versus in-plane lattice constant ratio of $\approx 1.3$ 
(Fig.\ref{fig:bco-tf-0}) [\onlinecite{cazorla17,cazorla18}]. The competing structures in BCO are the 
paraelectric (PE) orthorhombic ${\cal O}$ phase and the FE monoclinic ${\cal M}$ phase [\onlinecite{cazorla17}]. 
Both competitive phases have cells that are slightly distorted versions of the ideal cubic perovskite 
structure, with $c/a \approx 1$. The mentioned FE phases present spontaneous polarizations along quite 
different crystallographic directions, namely, pseudocubic $[001]_{\rm pc}$ for ${\cal T}$ and 
$\sim[111]_{\rm pc}$ for ${\cal M}$ (Fig.\ref{fig:bco-tf-0}). As regards magnetism, both the ${\cal O}$ 
and ${\cal M}$ phases display G-type AFM order (Fig.\ref{fig:bco-tf-0}) with a quite high N{\'e}el 
temperature $T_{\rm N} \sim 500$~K. The ${\cal T}$ phase, on the contrary, presents C-type AFM order 
(Fig.\ref{fig:bco-tf-0}) with a relatively low $T_{\rm N} \sim 310$~K [\onlinecite{cazorla17,cazorla18}].

The competition between BCO polymorphs is strongly affected by epitaxial strain, which is realized in practice
by growing epitaxially thin films on perovskite substrates, as this imposes an in-plane lattice constant 
$a_{\rm in}$ in the system. At zero-temperature conditions, a strain-driven ${\cal T} \to {\cal M}$ phase 
transformation has been predicted to occur at $a_{\rm in} = 3.925$~\AA~ that involves rotation of the 
polarization and the appearance of anti-phase oxygen octahedral rotations along the three pseudocubic 
directions [\onlinecite{cazorla18}]. The ${\cal O}$ phase remains close in energy to the ${\cal M}$ 
polymorph over the whole $a_{\rm in}$ interval but never becomes stable at zero temperature. Meanwhile, 
the N{\'e}el temperature of epitaxially grown BCO thin films decreases mildly with increasing $a_{\rm in}$
[\onlinecite{cazorla18}].

\begin{figure*}[t]
\centerline
        {\includegraphics[width=0.90\linewidth]{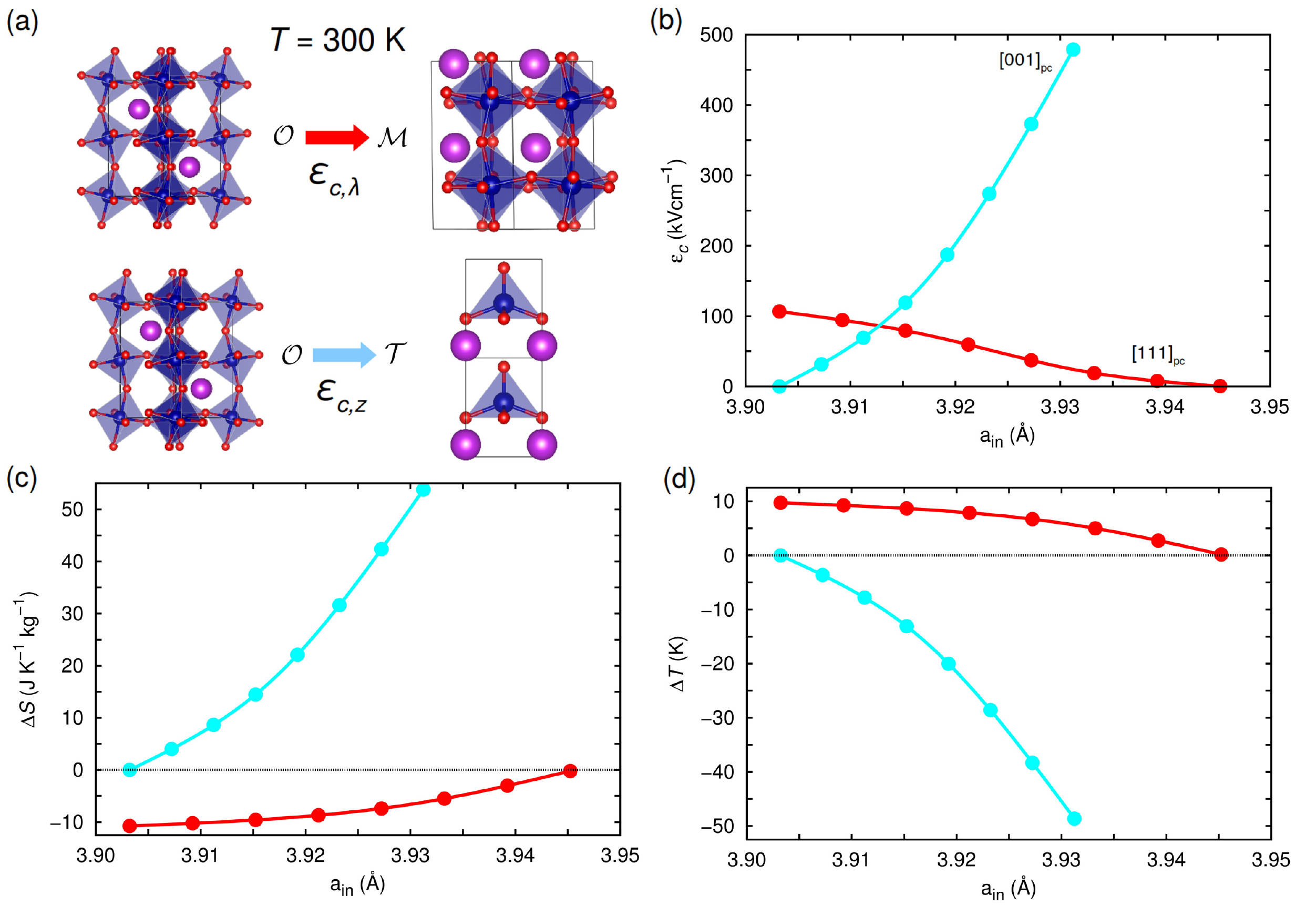}}
\caption{Direct (${\cal O} \to {\cal M}$, red) and inverse (${\cal O} \to {\cal T}$, blue) electrocaloric
         effects in $(100)$-oriented BCO thin films at room temperature estimated with first-principles 
         methods and the Clausius-Clapeyron approach. (a)~Sketch of the ${\cal E}$--induced phase 
         transformations. (b)~Critical electric field expressed as a function of in-plane lattice parameter;
         the two involved electric field orientations are indicated in pseudocubic Cartesian notation.
         (c)~Room-temperature entropy and (d)~adiabatic temperature shifts expressed as a function of in-plane
         lattice parameter. Reproduced from work [\onlinecite{cazorla18}].}
\label{fig:bco-tf-II}
\end{figure*}

Cazorla and ${\rm \acute{I}}$${\rm \tilde{n}}$iguez have performed first-principles Helmholtz 
free-energy calculations within the quasi-harmonic approximation for the three relevant BCO polymorphs to 
determine their relative stability as a function of $T$ and $a_{\rm in}$ (Sec.\ref{sec:clausius}) 
[\onlinecite{cazorla18}]. The predicted $T$--$a_{\rm in}$ phase diagram for epitaxially grown BCO thin films 
is shown in Fig.\ref{fig:bco-tf-I}a. For relatively small $a_{\rm in}$'s, it is found that the ${\cal T}$ 
phase dominates and extends its stability region to temperatures much higher than observed in bulk BCO [\onlinecite{cazorla17}]. 
The reason for this ${\cal T}$ stability enhancement is that the competing ${\cal O}$ polymorph remains highly 
strained at such $a_{\rm in}$ conditions, hence its free energy increases considerably as compared to the 
bulk case. As $a_{\rm in}$ is increased, the ${\cal T}$ phase eventually is replaced by the ${\cal O}$ 
and ${\cal M}$ polymorphs, yielding a very rich phase diagram that exhibits concurrent structural and 
spin-ordering transformations.

Figures~\ref{fig:bco-tf-I}b--d show the calculated Helmholtz free-energy differences $\Delta
\tilde{F}_{\rm harm}$ between the three relevant BCO polymorphs expressed as a function of $T$
and $a_{\rm in}$ [\onlinecite{cazorla18}]. Those calculations take into account all possible
sources of entropy, namely, magnetic and vibrational, and the interplay between spin disorder
and lattice vibrations. It is found that at high temperatures ($T \gtrsim 600$~K) the vibrational
contributions to $\tilde{F}_{\rm harm}$ always favor the ${\cal O}$ and ${\cal M}$ phases over
the ${\cal T}$ phase. Nevertheless, whenever a polymorph becomes magnetically disordered, the
corresponding Helmholtz free energy decreases significantly as a consequence of $T$--induced
lattice phonon softenings [\onlinecite{cazorla18}]. Accordingly, abrupt $\Delta \tilde{F}_{\rm harm}$
changes appear in Figs.\ref{fig:bco-tf-I}b--d at the corresponding AFM~$\to$~PM magnetic transition
temperatures. The strong spin-phonon couplings in epitaxially grown BCO thin films are responsible
for the stabilization of the ${\cal O}$--AFM(G) phase at temperatures near ambient and
$3.88$~\AA~$\le a_{\rm in} \le 3.96$~\AA. Indeed, when spin-phonon couplings are disregarded the
${\cal O}$ phase becomes stable only at high temperatures when reaches the paramagnetic state
(Figs.\ref{fig:bco-tf-I}b--d) [\onlinecite{cazorla18}].

The $a_{\rm in}$ region $3.89$~\AA~$\le a_{\rm in} \le 3.93$~\AA~ is particularly relevant from 
a practical perspective since many perovskite substrates present lattice constants in this range. 
Interestingly, a $T$--induced reentrant behavior that is reminiscent of bulk BCO under compression 
[\onlinecite{cazorla17}] occurs therein: upon heating, the BCO film transforms first from a FE 
(${\cal T}$--AFM(C) or ${\cal M}$--AMF(G)) phase to a PE (${\cal O}$--AFM(G)) state, then back to 
a FE (${\cal T}$--PM) phase, and finally to a PE (${\cal O}$--AFM(G)) state. Of particular interest 
is the PE ${\cal O}$--AFM(G) region appearing near room temperature $T_{\rm room}$, which is surrounded 
by two FE phase domains presenting markedly different features. In particular, the phase diagram in 
Fig.\ref{fig:bco-tf-I}a suggests that the ${\cal O}$ phase may be transformed into the ${\cal T}$ 
or ${\cal M}$ states by applying an electric field ${\cal E}$ along the [001]$_{\rm pc}$ or 
[111]$_{\rm pc}$ directions, respectively. Such ${\cal E}$--driven phase transformations involve 
drastic structural changes as well as magnetic transitions, hence big entropy changes are likely to 
occur as a consequence.

Figure~\ref{fig:bco-tf-II} shows the direct EC effect associated with the field-induced ${\cal O} 
\to {\cal M}$ transformation (Fig.\ref{fig:bco-tf-II}a), in which the entropy of the system decreases 
($\Delta S < 0$, Fig.\ref{fig:bco-tf-II}c). Since the critical temperature for the ferroelectric-paraelectric 
phase transition in BCO is close to $1,000$~K [\onlinecite{cazorla17}], thermal effects on the electric 
polarization can be disregarded near room temperature [i.e., $\Delta P (T) \approx \Delta P (0)$ in 
Eqs.(\ref{eq:clausiusS}) and (\ref{eq:clausiusE})]. For the smallest $a_{\rm in}$ values, a maximum 
adiabatic temperature change $\Delta T$ of $+10$~K (Fig.\ref{fig:bco-tf-II}d) is estimated for a maximum 
critical electric field of $110$~kVcm$^{-1}$ (Fig.\ref{fig:bco-tf-II}b). The magnitude of this effect 
and of the accompanying critical electric field decrease with increasing $a_{\rm in}$, as the region 
of ${\cal M}$ stability is approached. Similarly, Fig.\ref{fig:bco-tf-II} shows the inverse EC effect 
associated with the field-induced ${\cal O} \to {\cal T}$ transformation (Fig.\ref{fig:bco-tf-II}a), 
in which the entropy of the film increases ($\Delta S > 0$, Fig.\ref{fig:bco-tf-II}c). A maximum 
$\Delta T$ of $-50$~K (Fig.\ref{fig:bco-tf-II}d) is estimated for a maximum critical electric field of 
$500$~kVcm$^{-1}$ at $a_{\rm in} = 3.93$~\AA~ (Fig.\ref{fig:bco-tf-II}b). The magnitude of this effect
and of the corresponding ${\cal E}_{c}$ decrease with decreasing $a_{\rm in}$, as the region of ${\cal T}$ 
stability is approached. 

The predicted giant $\Delta T$ and $\Delta S$ values, which can be achieved with relatively small 
driving fields, turn epitaxially grown BCO thin films into very attractive EC materials. For instance, 
the coexistence of direct and inverse EC effects suggests a possible refrigeration cycle based on the 
direct transformation between the high-entropy FE ${\cal T}$ and the low-entropy FE ${\cal M}$ phases as 
induced by ${\cal E}$ rotation, with a cooling performance equal to the sum of the individual ${\cal O} 
\leftrightarrow {\cal T}$ and ${\cal O} \leftrightarrow {\cal M}$ cycles. Moreover, chemical substitution
appears to be an alternative strategy for stabilizing phases that are similar to the ${\cal M}$ (alike to 
the ground state of BiFeO$_{3}$ [\onlinecite{cazorla15}]) and ${\cal O}$ (the most common among perovskites 
[\onlinecite{peng18})] polymorphs discussed here, and to control the corresponding magnetic transition 
temperatures. One particular example is provided by BiCo$_{1-x}$Fe$_{x}$O$_{3}$ solid solutions, where a 
morphotropic transition between a ${\cal T}$--like and a ${\cal M}$--like phase is observed to occur at 
room temperature [\onlinecite{azuma08,azuma18}]. Likewise, bulk Bi$_{1-x}$La$_{x}$CoO$_{3}$ appears to 
be a good candidate where to realize field-driven ${\cal O} \to {\cal T}$ transformations 
[\onlinecite{cazorla17}]. Hence BCO offers a variety of experimental possibilities to achieve giant EC, 
bringing new exciting prospects to the field of solid-state cooling. 

\begin{figure*}[t]
\centerline
        {\includegraphics[width=0.90\linewidth]{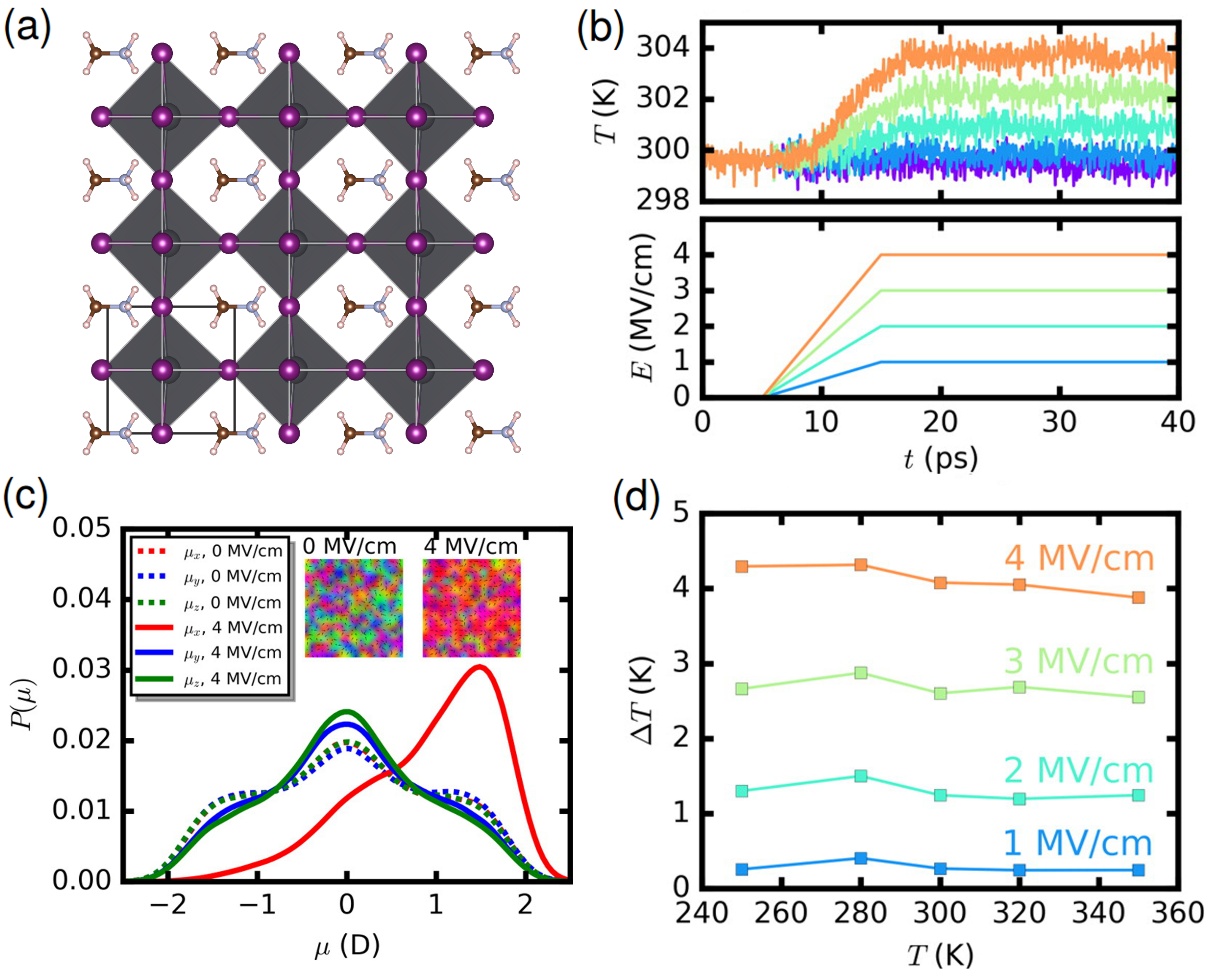}}
\caption{Electrocaloric effects in the hybrid perovskite CH$_{3}$NH$_{3}$PbI$_{3}$ predicted with
         molecular dynamics simulations and the direct method. (a)~Sketch of the atomic structure of 
         MAPbI$_{3}$. C, H, N, Pb, and I atoms are represented with brown, pink, violet, blue, and 
         purple spheres, respectively. (b)~Adiabatic thermal change of MAPbI$_{3}$ in response to 
         electric fields of different intensity. (c)~Probability distribution functions of the 
         molecular CH$_{3}$NH$_{3}$ dipoles under several electric fields. The insets show the 
         orientations of the molecular dipoles obtained from molecular dynamics simulations performed 
         at $300$~K. (d)~Temperature and electric field dependence of the electrocaloric effect in 
         MAPbI$_{3}$. Reproduced from work [\onlinecite{liu16}].}
\label{fig:mapi-ec}
\end{figure*}

\subsection{Bulk hybrid perovskite CH$_{3}$NH$_{3}$PbI$_{3}$}
\label{subsec:mapi}
Hybrid halide perovskites described with the general chemical formula AB$X_{3}$ ($X =$ F, Cl, Br, 
or I) consist of inorganic B$X_{6}$ octahedra, with B ions typically being Pb or Sn, and A organic 
molecular cations (e.g., CH$_{3}$NH$_{3}$, see Fig.\ref{fig:mapi-ec}a). The archetypal hybrid halide 
perovskite is CH$_{3}$NH$_{3}$PbI$_{3}$ (MAPbI$_{3}$), which has received a lot of attention in the 
past decade owing to its desirable solar matching optical bandgap, long carrier lifetime and diffusion 
length and some other remarkable functional properties [\onlinecite{yang17,tan14,ren20}]. 

Halide perovskites in general (e.g., HC(NH$_{2}$)$_{2}$PbBr$_{3}$ and MD--NH$_{4}$I$_{3}$ [\onlinecite{ye18}]) 
display elementary properties that are not observed in the analogous perovskite oxides, such as structural 
softness, lightweight, and low synthesis temperatures [\onlinecite{horiuchi08}]. On the other hand, 
the existence of ferroelectricity in halide perovskites remains a controversial and long-lasting topic 
of research. For instance, piezoelectric force microscopy and scanning electron microscopy studies have 
reported polar domains in tetragonal MAPbI$_{3}$, and weak ferroelectricity has been suggested by dielectric, 
piezoelectric and second harmonic generation measurements [\onlinecite{rakita17}]. However, traditional 
methods that identify ferroelectric insulators convincingly fail to characterize the polar properties of 
MAPbI$_{3}$ due to the technical difficulties ecountered in the distinction between ferroelectric and 
ferroelastic domains in this material [\onlinecite{hoque16}]. 

At temperatures slightly above ambient conditions, MAPbI$_{3}$ adopts a highly-symmetric cubic phase (space 
group $Pm\overline{3}m$) that is centrosymmetric and consequently cannot exhibit macroscopic electric 
polarization. Meanwhile, neutron powder diffraction experiments have shown that at room temperature the 
molecular cations in MAPbI$_{3}$, MA, are rotationally disordered [\onlinecite{weller15}]. Consistently, 
both \emph{ab initio} and classical molecular dynamics simulations have revealed fast reorientational
dynamics of the molecular cations with small relaxation times of $\sim 1$~ps near room temperature
[\onlinecite{mattoni15,mosconi14,carignano15}]. 

It is worth noticing that the MA cations in MAPbI$_{3}$ carry an intrinsic electric dipole since 
individually they do not fulfill inversion symmetry. Therefore, under the action of an external 
electric field the MA rotations can be partially frustrated and the molecular cations aligned, 
thus inducing some sort of polar ordering. Such a ${\cal E}$-induced MA ordering effect could potentially 
lead to large EC effects due to the concomitant reduction in the orientational entropy of the molecular
cations. Actually, colossal caloric effects near room temperature (i.e., $|\Delta T| \approx 50$~K) 
stemming from similar molecular ordering mechanisms, although induced by hydrostatic pressure, have been 
reported recently for plastic crystals with orientational disorder like neopentylglycol, 
(CH$_{3}$)$_{2}$C(CH$_{2}$OH)$_{2}$ [\onlinecite{cazorla19c,li19,lloveras19,lloveras20}].     

Liu and Cohen have employed MD simulations and classical force fields to investigate how molecular
ordering in MAPbI$_{3}$ responds to external electric fields, and what EC effects result from the accompanying
polar response [\onlinecite{liu16}]. In particular, the authors used a standard AMBER force field recently 
developed by Mattoni \emph{et al.} [\onlinecite{mattoni15}] to estimate directly (Sec.\ref{sec:direct}) 
room temperature EC effects induced by electric fields of intensities $0 < {\cal E} < 4$~MV/cm
(Fig.\ref{fig:mapi-ec}b). The MD simulations involved the use of large supercells containing 96,000 atoms, 
Nos\'{e}-Hoover thermostats, Parrinello-Rahman barostats, and simulation times longer than $2$~ns 
[\onlinecite{liu16}].  

Field-induced MA dipole alignment has been observed in Liu and Cohen's room-temperature MD simulations 
as shown by the shift in the peak of the probability distribution function estimated for the molecular 
dipoles calculated along the three Cartesian directions, $\lbrace \mu_{i} \rbrace_{x,y,z}$ 
(Fig.\ref{fig:mapi-ec}c). However, it has been found that even under relatively high electric fields of 
the order of $|{\cal E}| \sim 1$~MV/cm the MA cations still can rotate, thus the resulting electric 
polarization is still quite small (i.e., $P \sim 0.01$~C/m$^{2}$ [\onlinecite{stroppa15}]). This behaviour 
is in sharp contrast to what is observed for archetypal ferroelectrics like BaTiO$_{3}$ and PbTiO$_{3}$ 
where much larger electric dipole moments are achieved upon application of much smaller electric bias 
(i.e., $P \sim 0.1$~C/m$^{2}$ [\onlinecite{cazorla15}]). 

The adiabatic temperature changes predicted for MAPbI$_{3}$ at ambient conditions amount to $0.3$~K for an
electric field of $1$~MV/cm and to $4.1$~K for ${\cal E} = 4$~MV/cm (Fig.\ref{fig:mapi-ec}d) 
[\onlinecite{liu16}]. The value of these external electric fields are quite high and thus likely to cause 
leakage current issues in practice due to the small band gap of MAPbI$_{3}$. In addition, the estimated
adiabatic temperature shifts are quite modest in comparison to those observed in ferroelectric oxide 
perovskites, which may oscillate from few degrees up to $20$~K [\onlinecite{shirsath20}]. Nevertheless, the 
temperature dependence of the calculated $\Delta T$ around $T = 300$~K is very weak (Fig.\ref{fig:mapi-ec}d), 
which may be favorable for minimizing practical hysteresis and irreversibility problems 
[\onlinecite{cazorla19b,lloveras19,lloveras20}]. Such a $\Delta T$ temperature behaviour typically is found
in relaxor ferroelectrics (see next subsection), which hints at the structural similarities between 
MAPbI$_{3}$ and other materials that exhibit polar domains at the nanoscale and low temperatures 
[\onlinecite{acosta16}].

Liu and Cohen also have explored the combined action of epitaxial strain and electric fields on the 
polar and electrocaloric properties of MAPbI$_{3}$ at room temperature. In fact, the application of 
mechanical stresses offers very promising avenues in the field of caloric materials and solid-state
cooling [\onlinecite{sagotra17,sagotra18,min20}]. It has been found that under a small compressive 
biaxial strain of $2$\% a more complete alignment of the MA cations can be achieved for an electric 
bias of $1$~MV/cm (namely, a two-fold enhacement in comparison to the unstrained case [\onlinecite{liu16}]). 
Likewise, a significant increase on the electrocaloric response of MAPbI$_{3}$ has been demonstrated: 
for a compressive epitaxial strain of $4$\% the estimated $\Delta T$'s are about two or three times 
larger than those calculated for the analogous free-standing system (e.g., $5.2$~K for ${\cal E} = 
2$~MV/cm [\onlinecite{liu16}]). These MD results, although may seem not relevant from an applied 
point of view, illustrate very well the capabilities of atomistic simulations for analyzing and 
predicting complex dynamical effects that can lead to original thermal behaviour in functional 
materials. 

\begin{figure}[t]
\centerline
        {\includegraphics[width=1.00\linewidth]{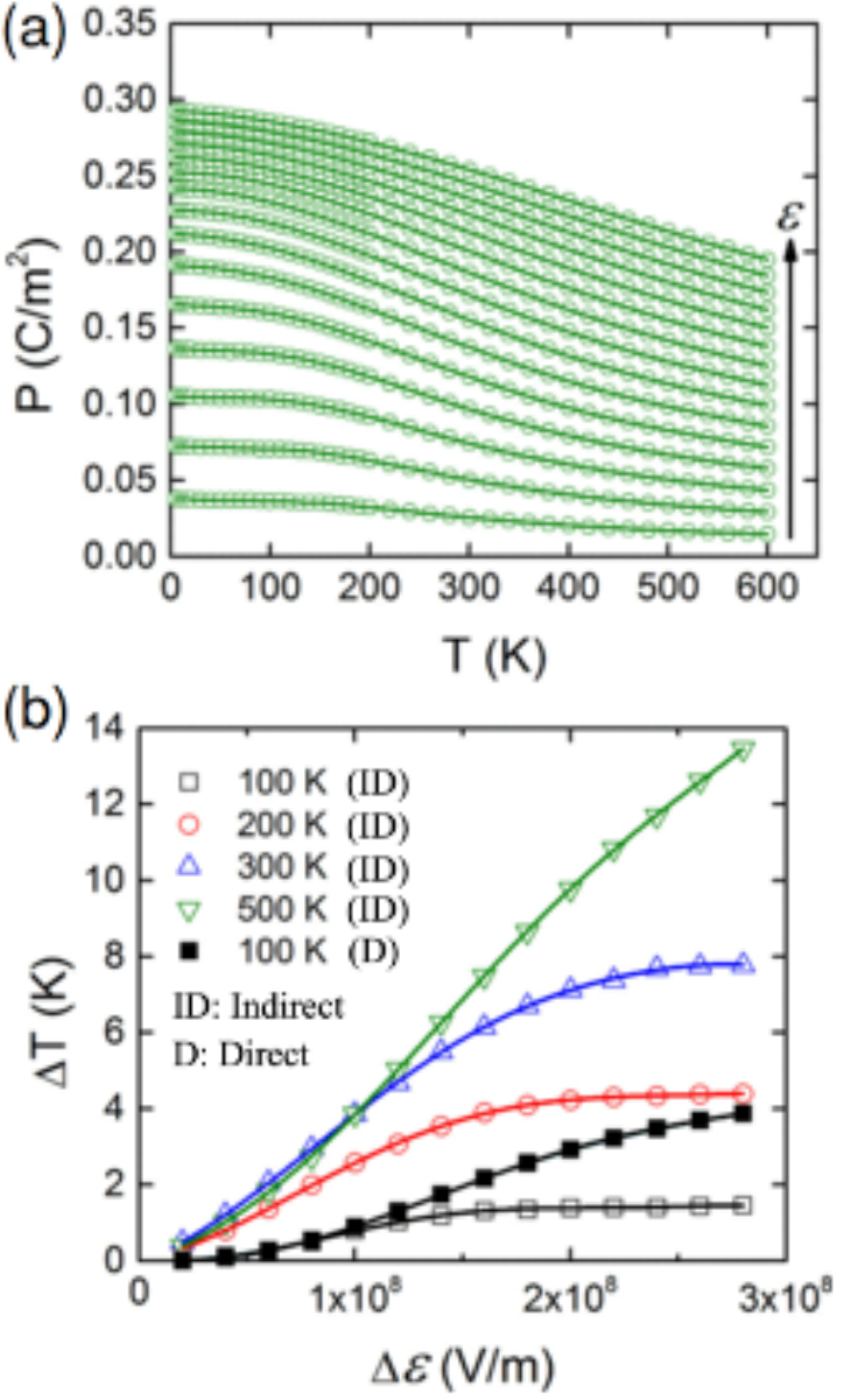}}
\caption{Ferroelectric and electrocaloric properties estimated for the relaxor BaZr$_{1-x}$Ti$_{x}$O$_{3}$
         by using an effective Hamiltonian model and molecular dynamics (direct method) and Monte Carlo 
         simulations (indirect method). (a)~Temperature depedence of the polarization under different
         electric bias. The electric field is applied on a fixed direction and its module changes
         from $2 \cdot 10^{7}$ to $3 \cdot 10^{8}$~V/m in increments of $2 \cdot 10^{7}$~V/m. (b)~The
         adiabatic temperature change associated with the electrocaloric effect expressed as a function of 
         temperature and maximum electric field. Reproduced from work [\onlinecite{jiang17}].}
\label{fig:relaxor-ec}
\end{figure}

\subsection{Relaxor ferroelectric BaZr$_{1-x}$Ti$_{x}$O$_{3}$}
\label{subsec:relaxor}
Relaxor ferroelectrics are a class of ferroelectric materials that present a dielectric response that
is frequency-dependent and broad in temperature while remain macroscopically paraelectric down to absolute 
zero temperature [\onlinecite{rozic11,pirc11,bokov12}]. Relaxor ferroelectrics typically exhibit high dielectric 
constants and also large electrostriction [\onlinecite{cross80}]. The microscopic origins of these phenomena 
are thought to be related to the existence of polar nanoregions below a particular temperature called the 
Burns temperature [\onlinecite{bokov06}]. Examples of archetypal relaxor ferroelectrics are lead magnesium 
niobate (Pb$_{3}$MgNb$_{2}$O$_{9}$ --PMN--), lead scandium niobate (PbSc$_{1-x}$Nb$_{x}$O$_{3}$ -- PSN --)
and solid solutions like barium titanate-bismuth zinc niobium tantalate (BT-BZNT) and barium titanate-barium 
strontium titanate (BT-BST) [\onlinecite{kong19,mohanty19,zhou16}].   
   
Relaxor ferroelectrics appear to be also promising electrocaloric materials in which original thermal
behaviour is driven by electric fields. For example, relatively large $\Delta T$'s of $2$--$3$~K have been 
measured directly in PMN-based oxides under small electric fields of $90$~kV/cm [\onlinecite{rozic11}]. The 
magnitude of the $\Delta T / {\cal E}$ coefficient is largest at the critical point in which the paraelectric
to ferroelectric phase transition changes from first-order type to second-order type. Another illustrative
example is provided by the lead-free relaxor ferroelectric Ba(Zr$_{0.8}$Ti$_{0.2}$)O$_{3}$ in which a 
large adiabatic temperature change of $\sim 5$~K has been measured over a broad temperature interval
of $30$~K [\onlinecite{qian14}].  

Jiang \emph{et al.} have used a first-principles based effective Hamiltonian along with classical 
molecular dynamics and Monte Carlo simulations to investigate the electrocaloric response of the 
relaxor ferroelectric Ba(Zr$_{0.5}$Ti$_{0.5}$)O$_{3}$ (BZT) in the temperature interval $100 \le T 
\le 500$~K (Fig.\ref{fig:relaxor-ec}) [\onlinecite{jiang17}]. The employed BZT effective Hamiltonian 
has been shown to reproduce successfully the characteristic temperatures measured in experiments 
as well as the existence and the dynamics of polar nanoregions [\onlinecite{akbarzadeh12}]. 
Jiang \emph{et al.}'s simulations involve very large simulation cells containing $\sim 10,000$ atoms 
and the use of the indirect and direct estimation techniques described in Secs.\ref{sec:indirect} 
and \ref{sec:direct}.  
 
Figure~\ref{fig:relaxor-ec} shows the estimated variation of the electric polarization in the BZT 
relaxor as a function of temperature and electric field module along with the corresponding 
electrocaloric adiabatic temperature changes. Relatively large $\Delta T$ of about $8$~K are predicted
near room temperature for moderately large electric fields of the order of $\sim 1$~MV/cm. The
temperature and electric field dependences of the estimated adiabatic temperature intervals are 
not monotonous, as it is expected from relaxor ferroelectrics [\onlinecite{zhou16}]. In particular, 
small values of the adiabatic electrocaloric coefficient, defined as $\left( \partial T / \partial 
{\cal E} \right)_{S}$, are obtained at low ${\cal E}$'s, followed by a sustained increase up to 
a maximum value beyond which it decreases for larger electric fields.   

Interestingly, Jiang \emph{et al.} have found that indirect and direct estimations of the electrocaloric
effect in BZT do not coincide at temperatures below the corresponding Burns point (Fig.\ref{fig:relaxor-ec}b) 
when polar nanoregions emerge [\onlinecite{jiang17}]. The reasons for such numerical inconsistencies are 
related to the occurrence of non-ergodic processes at low temperatures. In this case, direct estimation 
techniques can reproduce non-equilibrium phenomena with certain reliability and thus are the methods of 
choice for simulating caloric effects in relaxor ferroelectrics. Actually, a similar failure of indirect 
methods caused by nonergodicity has been demonstrated experimentally for the ferroelectric relaxor polymer 
poly(vinylidene fluoride-trifluoroethylene-chlorofluoroethylene) [P(VDF-TrFE-CFE)] [\onlinecite{lu10}]. 
The present study illustrates once again the capabilities of first-principles based methods for simulating 
complex atomistic behaviour in promising materials that render sizable electrocaloric effects.

\section{Conclusions}
First-principles based computational techniques for simulation of electrocaloric effects in oxide 
perovskites and similar polar materials are already madure and well established. These simulation 
methods (e.g., quasi-harmonic free-energy methods and effective Hamiltonians) were originally 
developed for the study of temperature and field-induced phase transformation in materials and have
already demonstrated great success in other research disciplines. 

The suite of computational approaches reviewed in this Chapter can be used to reproduce with reliability 
the electrocaloric performance of complex materials like multiferroics, in which the electronic and 
lattice degrees of freedom are strongly coupled, organic-inorganic halide perovskites, in which the 
molecular cations can be orientationally disordered, and ferroelectric relaxors, in which the existence 
of polar nanoregions and non-ergodic processes underpin their physical behaviour. Analogous electrocaloric 
simulation success can be achieved for other functional materials that exhibit complex and unconventional 
responses to electric fields.        

First-principles based simulation of electrocaloric effects, therefore, can help enormously in 
developing new materials and strategies for boosting solid-state cooling engineered on electric 
fields. Thus, the current pressing challenge of finding new refrigeration technologies that are 
environmentally friendly and highly scalable in size can benefit vastly from the outcomes of such 
accurate and reliable simulation methods.

\end{document}